\documentclass{article}

\usepackage{PRIMEarxiv}
\usepackage{textcomp}
\usepackage{caption}
\usepackage[ruled,vlined]{algorithm2e}
\usepackage[utf8]{inputenc} 
\usepackage[T1]{fontenc}    
\usepackage{hyperref}       
\usepackage{url}            
\usepackage{booktabs}       
\usepackage{amsfonts}       
\usepackage{nicefrac}       
\usepackage{microtype}      
\usepackage{lipsum}
\usepackage{fancyhdr}       
\usepackage{graphicx}       
\graphicspath{{media/}}     
\usepackage{amsmath,amssymb,amsfonts}
\title{Intent-driven scheduling of backup jobs
\thanks{\textit{\underline{Citation}}: 
\textbf{Dutta S., Brahmaroutu S.,. Intent-driven scheduling of backup jobs, Pages.... DOI:xxxxxx/yyyy.}} 
}

\author{
  Souvik Dutta ~and~ Suri Brahmaroutu \\
  Veritas Technologies, LLC. \\
  Santa Clara CA, United States of America\\
  \texttt{\{souvik.dutta, suri.brahmaroutu\}@veritas.com} \\
}
\begin{document}
\fontfamily{qtm}\selectfont

\maketitle

\begin{abstract}
Job scheduling under various constraints to achieve global optimization is a well-studied problem. However, in scenarios that involve time-dependent constraints, such as scheduling backup jobs, achieving global optimization may not always be desirable. This paper presents a framework for scheduling new backup jobs in the presence of existing job schedules, focusing on satisfying intent-based constraints without disrupting current schedules. 
The proposed method accommodates various scheduling intents and constraints, and its effectiveness is validated through extensive testing against a variety of backup scenarios on real-world data from Veritas Netbackup\textregistered ~customer policies.
\end{abstract}


\section{Introduction}

\begin{figure}[ht!]
\centering
\includegraphics[scale=0.45]{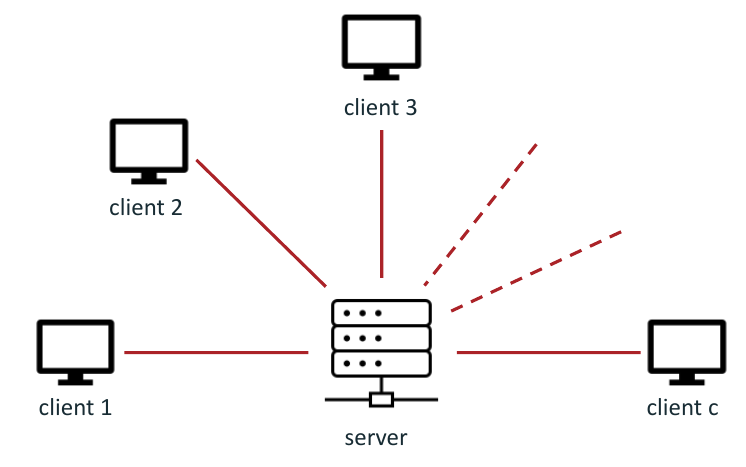}
\includegraphics[scale=0.35]{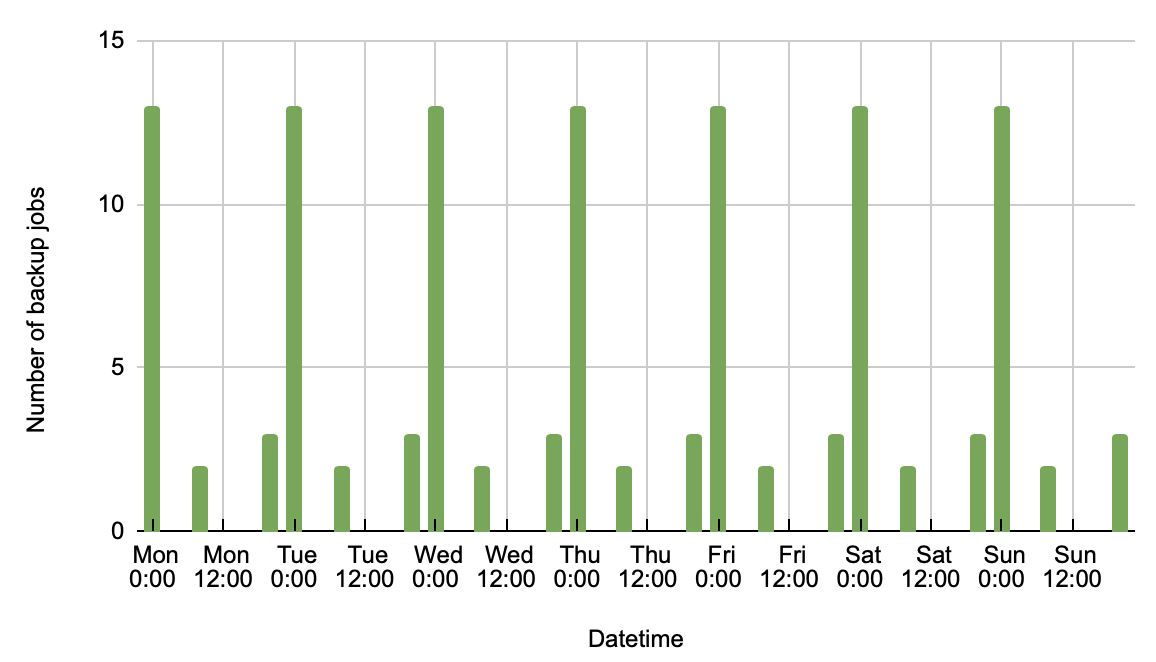}
\caption{\small \textbf{Left:} Backup client-server topology with $c$ clients; if the number of simultaneous backups is more than the concurrency limit of the server, backup jobs for some clients may fail. \textbf{Right:} Backup schedule for a customer, where 72\% backup jobs are clustered around midnight (from a Veritas customer's Netbackup policy configuration.}
\label{fig_topology}
\end{figure}

\hspace{5 mm}In the data backup and recovery domain, scheduling backup operations to minimize disruption to standard business processes is critical. However, as highlighted in \cite{b1}, users tend to schedule backups at specific hours of the day or week, leading to server overload during peak times and extended periods of inactivity otherwise (see Fig. \ref{fig_topology}). This clustering of backup jobs often exceeds server concurrency limits, resulting in a significant number of job failures. Indeed, \cite{b4} attributes the majority of these failures to inefficient scheduling practices. Further complicating this issue is the dynamic nature of server workloads, which must accommodate a fluctuating number of clients (physical and virtual machines). These challenges underscore the need to shift from user-defined scheduling to approaches that align with user \emph{intent}.

\hspace{5 mm}The objective of our research is two-fold:
\begin{itemize}
    \item rigorously define the constraints inherent in scheduling backup jobs, and
    \item enable backup administrators to incrementally develop backup policies that accurately reflect strategic \emph{intent}.
\end{itemize}
In \S \ref{sec:intentdef}, we explain why scheduling backup jobs needs to be handled differently from standard constrained optimization problems. In \S \ref{sec:constraints}, we formalize the problem statement and systematically list implicit and explicit constraints stemming from the user intent. In \S \ref{sec:algorithm}, we present the scheduling algorithm and explain the rationale for the prescription. Finally, in \S \ref{sec:results}, we evaluate the effectiveness of our approach by applying it to backup schedules for customers with PIIs redacted.

\section{User-intent for scheduling backup jobs}
\label{sec:intentdef}
\hspace{5 mm}Scheduling jobs is a well-studied problem in computer science and operations research, often approached with methods that prioritize jobs based on some metric such as importance (e.g., priority scheduling), queued time (e.g., first-come-first-served), or expected duration (e.g., shortest-job-first). These approaches work well for optimizing the order of independent jobs but fall short in enterprise backup scheduling, where an initial queue of jobs is typically predefined, and new jobs must be integrated with minimal disruption to existing schedules. Global optimization techniques, such as genetic algorithms, simulated annealing, or integer programming, aim to optimize the schedule globally, but may lead to an undesirable reconfiguration of pre-existing schedules. Furthermore, backup administrators impose domain-specific constraints (outlined in \S\ref{sec:constraints}), such as maintaining affinity between related schedules or limiting overlaps between consecutive backups. These requirements demand specialized approaches that maximally preserve schedule integrity while adapting to dynamic workloads.
    \begin{figure}[htbp]
    \centerline{\includegraphics[scale=0.7]{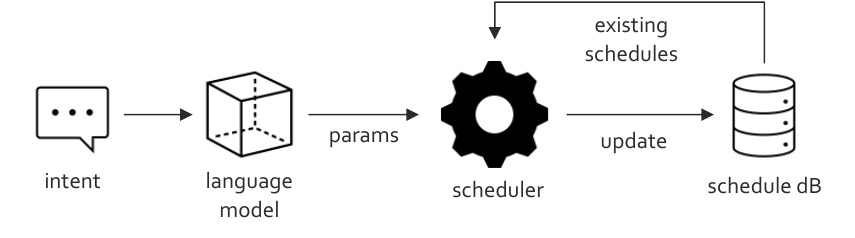}}
    \caption{From the intent, a small language model extracts parameters for the scheduler, which in turn integrates with the existing job windows, to create a new schedule that is designed to match the user intent.}
    \label{fig_goal}
    \end{figure}

\hspace{5 mm}In this paper, we propose an algorithm that translates the intent and constraints specified by the backup admin into a schedule that minimally disrupts existing jobs. For example, an administrator might specify: ``\emph{Backup asset VM16as\_v1 four times per week with minimal overlap with other backup jobs, and no more than twice on any day.}" or ``\emph{Backup VM16as\_v1 during lull periods for normal jobs.}" As in Fig. \ref{fig_goal}, this intent is processed through a language model that extracts key scheduling parameters required for scheduling, such as the number of new jobs requested, time period over which they will be distributed, average intended overlap, etc. The scheduling algorithm then relates these parameters with the distribution of existing jobs to generate a new schedule that best aligns with the specified intent.

\section{Problem set-up and constraints}
\label{sec:constraints}

\hspace{5 mm}Assume we have a topology with $c$ clients connected to a single backup server as in Fig. \ref{fig_topology} L. Over a length of time, these clients are being backed up via $n \geq c$ jobs. Each job is assigned a time-window $w_i$ of length $\delta_i$, centered at $\tau_i$:
\begin{equation}\label{eq:windows}
    w_i = (\tau_i - \delta_i/2, \tau_i + \delta_i/2), \qquad 
\end{equation}
within which the backup must be completed -- an example is shown in Table \ref{tab:scheds}.
\begin{table}[ht!]
\begin{center}
\setlength{\tabcolsep}{6pt} 
\renewcommand{\arraystretch}{1.5} 
\begin{tabular}{c|c|c|c||c|c}
Job window & Client & Start Time & End Time & $\tau$ & $\delta$ \\
\hline
$w_1$ & client 1  &  M 13:00  &  M 17:00  &  M 15:00    &  2.00   \\
$w_2$& &  W 15:00  &  W 18:00  &  W 16:30    &  1.50   \\
 \hline
$w_3$ & client 2  &  Su 23:00  &  M 00:30  &  Su 23:45    &  0.75   \\       
\end{tabular}
\end{center}
\captionof{table}{Example schedules for 2 clients and 3 jobs.}\label{tab:scheds}
\end{table}

\hspace{5 mm}We want to schedule $k \in \mathbb Z^+$ new non-overlapping windows $\{w_{n+1}, \ldots w_{n+k}\}$ for an incoming client subject to both implicit and explicit (user-defined) constraints. Implicit constraints may include hardware limitations (such as the server's capacity to process a maximum number of concurrent jobs) or the periodic nature of all jobs within a week. Explicit, user-defined constraints can involve requirements like a minimum spacing between consecutive jobs for the same client or the need for affinity with existing schedules. 
\vspace{1 mm}

\hspace{5 mm}In this paper, we assume that the window width $\delta_i$ is fixed as $\Delta > 0$ for all $i \in [1,n+k]$ only for simplicity of notation. However, the logic is adaptable to arbitary $\delta_i$.
\subsection{Implicit constraints}
\begin{enumerate}
    \item \emph{Concurrency limit}: Let $f(n;t)$ be the number of existing jobs ($\leq n$) at time $t$ and $\mathcal J$ be the maximum number of concurrent jobs the backup server can process. Then, 
    \begin{equation}
       0 \leq f(n;t) \leq \mathcal J, \qquad \forall n, t.
    \end{equation}
    \item \emph{Disjointedness}: Unless backup jobs have long durations, the $k$ new backup windows must be mutually disjoint, meaning that $w_{n+1} \cap w_{n+2} = \phi$, $w_{n+2} \cap w_{n+3} = \phi$, and so on. Alternately,
    \begin{equation}\label{eq:disjoint}
        \tau_{n+i+1} - \tau_{n+i} \geq \Delta, \qquad \forall i \in \{1, \ldots, k-1\}.
    \end{equation}
    \item \emph{Periodicity}: Enterprise backup schedules are usually
periodic in time, that is:
    \begin{equation}\label{eq:f_periodic}
        f(n;t+P) = f(n;t), \qquad P > 0.
    \end{equation}
    This requirement closes the circle on \eqref{eq:disjoint} as $(\tau_{n+1} - \tau_{n+k})(\text{mod }P) \geq \Delta$.
\end{enumerate}

\subsection{Explicit constraints}
\begin{enumerate}
    \item \emph{Min spacing:} The backup admin may adjust the spacing in \eqref{eq:disjoint} to:
\begin{equation}\label{eq:spacing}
    \tau_{n+2} - \tau_{n+1} \geq \epsilon, \qquad \epsilon > \Delta.
\end{equation}
This may be necessary to minimize cloud costs from performing backups, write costs into storage, and the user's recovery point objective requirements.
\item \emph{No. of windows:} The admin selects $k$, the number of backup windows required for the new client. Generally, increasing 
$k$ adds more constraints to the scheduling problem. To ensure the problem is well-defined without violating \eqref{eq:spacing}, we require
\begin{equation}\label{eq:sanity}
    k \times \epsilon < P.
\end{equation}
\item \emph{Expected overlap}: The admin may choose an expected overlap $\alpha \in [0, 1]$ with existing windows. For instance, if $\alpha \sim 1$, the algorithm will attempt to schedule new jobs highly correlated with the time windows of existing jobs. Conversely, when $\alpha \sim 0$, the new jobs will be staggered relative to the existing schedule. Also, this might be useful when the existing schedules are not for backups but for other high-priority jobs which must not be affected.
\item \emph{Self-affinity}: The admin may want to reinforce an affinity towards or away from newly scheduled windows with a parameter $0 \leq \omega \leq 1$. This parameter may help in breaking ties when several windows are equally likely to be suggested. 
\end{enumerate}
\vspace{1mm}

\hspace{5 mm}In the intent example above (reproduced here): ``\emph{Backup asset VM16as\_v1 four times per week with minimal overlap with other backup jobs, and no more than twice on any day.}", $\epsilon = 12$ hrs, $P = 24\times 7 = 168$ hrs, $k=4$, $\alpha = 0$. We note that constraint in \eqref{eq:sanity} is respected here as $4\times 12 < 168$, and so this is a well-posed problem. However, whether $\alpha$ is achievable depends on the existing schedules.

\section{Scheduling algorithm}
\label{sec:algorithm}
We begin by presenting the algorithm's pseudo-code, followed by a detailed explanation of each step.

\begin{algorithm}[H]
\caption{Pseudo-code for sampling algorithm}
\KwIn{Existing schedules $f(n; t)$ and intent parameters \{$\alpha$, $k$, $\epsilon$, $P$, $\omega$\}}
\KwOut{$\{\tau_i\}$ (set of scheduled time points)}
\vspace{2 mm}

Obtain the kernel density estimate $F_h(t)$ based on $f(n; t)$\;
Use $F_h(t)$ to create sampling distribution $G(\alpha, t)$\;

Initialize an empty set of sampled times: $T \gets \{\}$\;

\For{$i \gets 1$ \KwTo $k$}{
    \If{$G(\alpha,t) = 0 \; \forall t$}{
        \textbf{terminate with error:} ``Unable to proceed''\;
    }
    Sample $\tau_i$ from $G(\alpha,t)$\;
    Update $G(\alpha,t)$ to account for $\tau_i$\;
    $T \gets T \cup \{\tau_i\}$\;
}

\Return $T$\;
\end{algorithm}

\subsection{Construct the kernel density estimate $F_h(t)$}
\hspace{5 mm}The core idea of this algorithm is to identify high and low-density regions of existing jobs over time, enabling the determination of suitable regions for sampling.  The kernel density estimate (KDE) provides an approximation of the underlying distribution from which $\{\tau_1, \ldots, \tau_n\}$ were drawn. KDEs have been previously used in the literature for cloud-resource configuration \cite{d1} and for serving cost-effective LLMs \cite{h1}. Let $F_h(t)$ be the KDE for $f(n;t)$ with bandwidth $h$. With an appropriate choice of $h$,\footnote{For the bandwidth choice, one can use the Silverman's rule of thumb \cite{b2} or Scott's rule \cite{b3}.} $F_h(t)$ reveals the temporal density profile of existing jobs. 

\begin{figure}[ht!]
    \centerline{\includegraphics[scale=0.55]{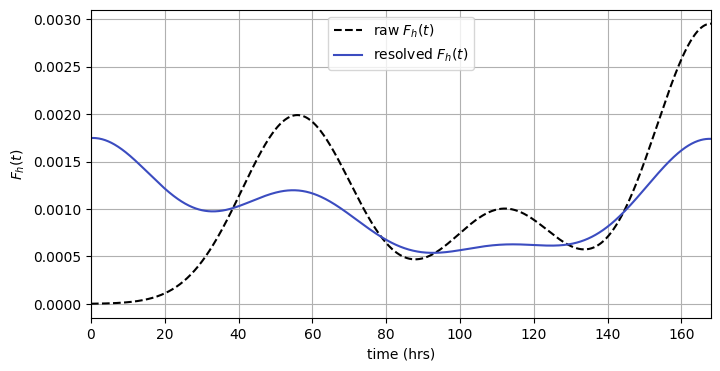}}
    \caption{Sample $F_h(t)$ with (solid blue) and without (dashed black) resolving edge effects and periodicity. The time axis is hours starting Mon 12:00 PM.}
    \label{fig_periodic}
    \end{figure}
    
\hspace{5 mm}One limitation of the bare $F_h(t)$ obtained using a library like \texttt{scipy.stats.gaussian\_kde} is the presence of edge-effects. While the true domain of $f(n;t)$ is bounded within $[0,P]$, the KDE inherently assumes an unbounded domain, extending over $(-\infty,\infty)$. This assumption introduces inaccuracies at the boundaries of $t=0^+$ and $t=P^-$ as the periodic nature of $f(n;t)$, as defined in \eqref{eq:f_periodic}, is not inherently captured. In other words, we need:
\begin{equation}\label{periodicity}
    F_h(0^+) \approx F_h(P^-), \qquad \int_0^P F_h(t) = 1.
\end{equation}

\hspace{5 mm}The edge effects and periodicity constraints can be simultaneously addressed through a domain expansion approach --  we expand the domain of $f(n;t)$ from $[0, P] \to [-u, P+u]$, where $u$ is comparable to $P$, e.g., $u= P/4$. The behavior of $f(n;t)$ outside its original domain is augmented as a ``wrap-around":
\begin{equation}
    f(n;t<0) = f(n;P-u\leq t \leq P), \qquad \text{and} \qquad f(n;t>P) = f(n;0\leq t \leq u).
\end{equation}

The KDE of $f(n;t)$ defined on the expanded domain is subsequently restricted to the interval $[0, P]$ and then normalized, to produce a PDF fully supported on $[0, P]$ and satisfying \eqref{periodicity}. In Fig. \ref{fig_periodic}, we show the raw $F_h(t)$ and the resolved $F_h(t)$ for a sample schedule.

\subsection{Compute the sampling distribution $G(\alpha, t)$}

\hspace{5 mm}Now, we want to construct a distribution $G(\alpha, t)$ that shows the following limiting behaviors:
\begin{itemize}
    \item At $\alpha \to 1$, $G(\alpha, t) = F_h(t)$,
    \item At $\alpha \to 0$, the minima of $G(\alpha, t)$ coincide with the maxima of $F_h(t)$.
\end{itemize}

We can achieve this schematically as:
\begin{equation}
    G(\alpha, t) = \alpha F_h(t) + (1-\alpha) F_h^{-1}(t),
\end{equation}
where we define $F_h^{-1}(t)$ is a function with an inverse density from that of $F_h(t)$. While there are multiple approaches to construct $F_h^{-1}(t)$, in this paper, we adopt a straightforward choice: $F_h^{-1}(t) = - F_h(t)$. As $F_h^{-1}(t) < 0$, by definition, it does not satisfy the non-negativity requirement of a valid probability density function (PDF); however, we can make $G(\alpha,t) \geq 0$ by a shift:
\begin{equation}\label{Gdef}
    G(\alpha, t) = (2\alpha-1) F_h(t), \qquad G(\alpha, t) \to G(\alpha, t) - \underset{t\in P}{\min}~G(\alpha, t)
\end{equation}
    \begin{figure}[htbp]
    \centerline{\includegraphics[scale=0.55]{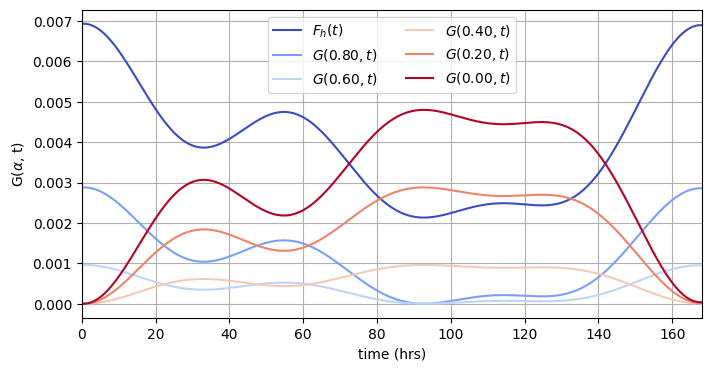}}
    \caption{The original KDE $F_h(t)$ and the distributions $G(\alpha, t)$ for various values of the overlap parameter $\alpha \in [0, 1]$. The time axis is hours starting Mon 12:00 PM.}
    \label{fig_Ght}
    \end{figure}
    
We can further normalize $G(\alpha, t)$ to convert this into a valid PDF -- we will not do it here. As can be deduced from \eqref{Gdef} or from Fig. \ref{fig_Ght}, the limiting conditions on $G(\alpha, t)$ at $\alpha \to 0$ and $\alpha \to 1$ are satisfied.

\subsection{Sampling from and updating $G(\alpha, t)$}
\begin{figure}[htbp]
    \centerline{\includegraphics[scale=0.55]{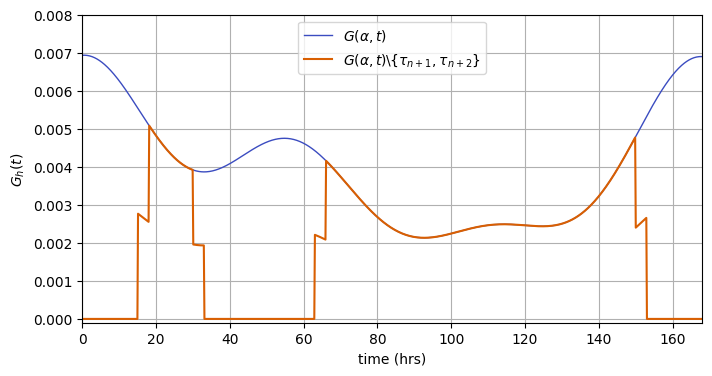}}
    \caption{Example evolution of $G(\alpha=1,t)$ after two sampling rounds with $\Delta = 15$ and reinforcement $\omega = 0.2$. The time axis is hours starting Mon 12:00 PM.}
    \label{fig_exclusion}
    \end{figure}

Given $G(\alpha,t)$, our objective is to sample $\{\tau_{n+1},\ldots,\tau_{n+k}\}$ such that the scheduling constraints are satisfied. To achieve this, we adopt a greedy approach iteratively:
\begin{enumerate}
    \item Identify $\tau_i = \underset{t\in P}{\text{argmax }} G(\alpha, t)$ with ties being broken randomly.
    \item Update $G(\alpha,t)$ by excluding a $2\Delta$-sized window $\mathcal X_i$ around $\tau_i$ such that 
    \begin{equation}\label{eq:exclude}
        G(\alpha,t \in \mathcal X_i) = 0, \qquad \mathcal X_i = [\tau_i-\Delta, \tau_i + \Delta]
    \end{equation} If the admin chooses an $\epsilon$ to replace $\Delta$ by \eqref{eq:spacing}, update \eqref{eq:exclude} with $\Delta \to \epsilon$. If the self-affinity parameter $\omega \neq 0$, then in a collar region of width $\Delta(1-\omega)$ around $\mathcal X_i$, $G(\alpha,t) \to G(\alpha,t)/2$.
\end{enumerate}
In Fig \ref{fig_exclusion}, we show the sampling distribution (not a PDF) after two sampling rounds. Provided that the condition in \eqref{eq:sanity} is not violated, $G(\alpha,t) \neq 0$ at least for some $t\in P$, ensuring the completion of the process.

\section{Test results and summary}
\label{sec:results}
We utilized the massive corpus of backup schedules across hundreds of Veritas customers (PII redacted) over 26,142 clients (physical and virtual machines) between Sep 1$^\text{st}$ and Sep 26$^\text{th}$ 2024, and and concluded the following:
\begin{itemize}
    \item A significant 83\% of backup jobs were configured using the \emph{default} settings of NetBackup policies, highlighting the difficulty associated with manually aligning policy configurations with user intent. 
    \item Approximately 3,400 backup jobs per week, across all Veritas Alta customers, failed with \texttt{exit status 196} (backup window closed), indicating a considerable number of failures that could potentially be mitigated through more optimized scheduling strategies. 
\end{itemize}

\begin{figure}[ht!]
\centering
\includegraphics[scale=0.45]{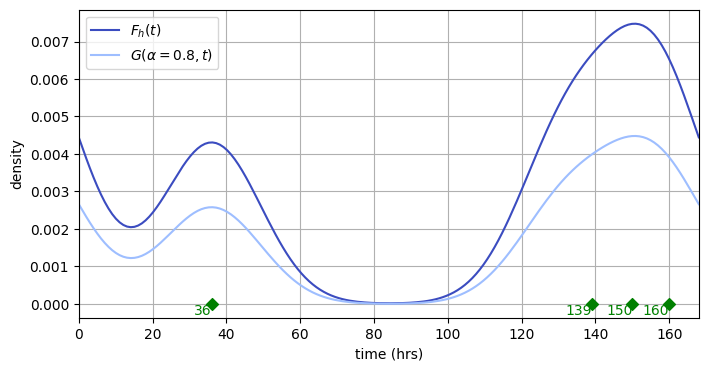}
\includegraphics[scale=0.45]{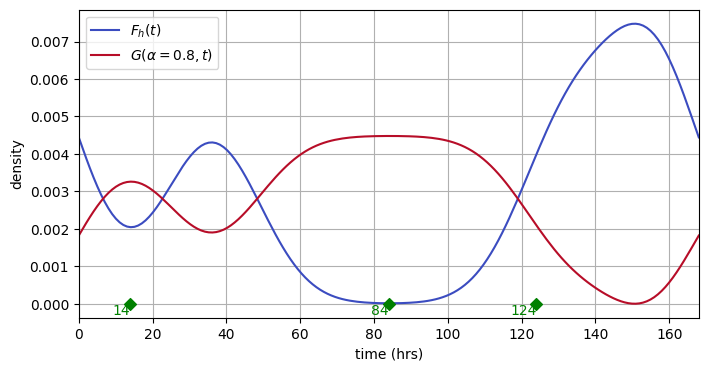}
\caption{\small \textbf{Left:} Recommended schedule for $k=4$ new jobs with minimum spacing $\epsilon=10$ hrs and expected overlap $\alpha = 0.8$. \textbf{Right:} Recommended schedule for $k=3$ new jobs with $\epsilon=40$ hrs and $\alpha = 0.2$. The time axis is hours starting Mon 12:00 PM.}
\label{fig:results}
\end{figure}

In Fig. \ref{fig:results}, we present the KDE for the existing backup job schedules on a single NetBackup primary server, over the course of a week. Incremental backup jobs were scheduled between 23:00-2:00 on Tuesdays and Saturdays, and full backup jobs were scheduled 23:00-5:00 on Sundays. On the left panel are the recommended backup times $\{\tau_1, \ldots \tau_4\}$ for the prompt ``\emph{Incremental backup for asset  VM16as\_v1 4 times with moderate overlap with my existing schedules. Ensure that they are spread apart by at least 10 hours}". 
\vspace{1 mm}

\hspace{5 mm}On the right panel is the output for the prompt ``\emph{I need to backup VM16as\_v1 3 times but try to schedule them when no other backups are happening, and not more frequently than once every 40 hours.}"

\section{Conclusions}
\hspace{5 mm}
In this paper, we have proposed a novel framework for scheduling backup jobs in environments that require a careful balance between hardware constraints, minimizing disruptions to existing operations, and aligning with user-defined intents. We highlighted the differences between traditional constrained optimization problems and the specific challenges posed by backup job scheduling. From existing job schedules and by parsing the user intent as parameters, we carefully constructed a distribution from which to iteratively sample future backup windows, so as to not disrupt ongoing operations (whether backups or other higher priority jobs).
\vspace{1 mm}

\hspace{5 mm}The proposed scheduling method was evaluated through extensive testing on thousands of Veritas Netbackup (PII redacted) customer backup policies with diverse potential intents. Our results confirm that, in contrast to traditional global optimization approaches, a more tailored, intent-driven scheduling framework can significantly improve the reliability of backup job management and minimize failures in complex, time-sensitive environments. 

\section*{Acknowledgments}
This was completed with support from the Data Protection Group at Veritas Technologies, LLC. SD would like to thank Thumbeti Sivaramaiah for useful discussions, and Liam Mcnerney and Atul Khandelwal for compiling field statistics.

\bibliographystyle{unsrt}  
\bibliography{draft}

\end{document}